\documentclass[aps,prb,twocolumn,10pt]{revtex4-1}
\usepackage{graphics,float}
\usepackage{amssymb,amsmath,amsfonts,latexsym}
\usepackage{bm,array}
\usepackage{psfrag}
\usepackage{epsfig}
\usepackage[caption=false]{subfig} 
\usepackage{slashed}
\floatstyle{plaintop}
\restylefloat{table}
\usepackage[flushleft]{threeparttable}
\usepackage{enumitem,booktabs}
\usepackage[referable]{threeparttablex}
\usepackage{longtable}
\renewlist{tablenotes}{enumerate}{1}
\makeatletter
\setlist[tablenotes]{label=\tnote{\alph*},ref=\alph*,itemsep=\z@,topsep=\z@skip,
partopsep=\z@skip,parsep=\z@,itemindent=\z@,labelindent=\tabcolsep,labelsep=.2em,leftmargin=*,align=left,before={\footnotesize}}
\makeatother
\usepackage{multirow}

\begin{document}

\title{Solution of the 1d Schr\"odinger Equation for a Symmetric Well}

\author{Lindomar Bomfim de Carvalho, Wytler Cordeiro dos Santos, and Eberth de Almeida Correa} 
\affiliation{Faculdade UnB Gama, Universidade de Bras\'ilia - UnB, 72444-240, Gama-DF, Brazil.}

\date{\today}

\begin{abstract}
We suggest a mathematical potential well with spherical symmetry and apply to the 1d Schr\"odinger equation. We use some well-known techniques as Stationary Perturbation Theory and WKB to gain insight into the solutions and compare them to each other. Finally, we solve the 1d Schr\"odinger equation using a numerical approach with the so-called Numerov technique for comparison. It can be a good exercise for undergrad students to grasp the above-cited techniques in a quantum mechanics course.          
\end{abstract}

\keywords{Potential Well; WKB; Numerov}

\maketitle

\section{Introduction}

In Quantum Mechanics books we usually find trivial examples when Stationary Perturbation Theory (SPT), Wentzel-Kramers-Brillouin (WKB) and even other techniques are discussed. The interesting applications are left to some complicated exercises at the end of the chapter\cite{Weinberg, Sakurai, Mahon, Griffths, Schiff, Cohen}. Most of the books apply those techniques to the simple harmonic oscillator with $V(x)\sim x^2$ or at most to the $x^4$ potential in 1d. Increasingly, the computer is becoming part of the physics courses, and it would be interesting to have certain classes of problems to be solved in a Quantum Mechanical course. 
 
Nowadays, several numerical techniques have been successfully employed to solve the Schr\"odinger equation to obtain both the energy levels and the respective wave functions. In particular, for this kind of differential equation, a powerful method among others is that proposed by Boris Vasil'evich Numerov\cite{Hairer}. This method takes advantage of the fact that the Schr\"odinger equation is an eigenvalue equation to handle the wave function subjected to certain boundary conditions in order to minimize the energy for each level. Its algorithm is very efficient and converges very fast with at least $O(h^6)$ of precision. Under this perspective, it is a good exercise to solve the 1d-Schr\"odinger equation using this technique to gain good insights about the Schr\"odinger equation during the classes\cite{Caruso}.      
Because of that, we suggest a mathematical potential well, which can be expanded in even exponent power series inside the well, becoming an excellent exercise to treat it perturbatively or using the WKB method or other well-known Quantum Mechanics techniques to compare with the powerful numerical results. In what follows we compare some states order by order to a good numerical calculation in determining the solutions of the 1d-Schr\"odinger equation. 

This article is sketched as follows: in section \ref{pot_well} we present the mathematical potential well. In section \ref{1d_schrod} we apply the potential well to the 1d-Schr\"odinger equation and analyses some particular aspects for both analytical and WKB approximation. In section \ref{SPT} we present the Stationary Perturbation Theory to calculate the energy levels and some wave functions for comparison. In section \ref{Num_Met} we discuss about the Numerov's numerical approach to solve the 1d-Schr\"odinger equation. Finally, in section \ref{conclu} we draw conclusions and perspectives. 

\section{The potential well}\label{pot_well}

Now we will introduce a mathematical potential whose series expansion is interesting because its symmetry and divergence at specific points. In 3D this mathematical potential is given by

\begin{eqnarray}
V(\mathbf{r})=V_0\left(\frac{1-kr\cot(kr)}{(kr)^2}\right)\label{pot3D}
\end{eqnarray} 

\noindent where $k$ has m$^{-1}$ units,  $r=\sqrt{x^2+y^2+z^2}$ in m, and $V_0$ in J. $k$ must be such that when $r$ equals some value, say $a$, the product $kr$ must be equals $\pi$. As a consequence $k=\pi/a$ and $V(\mathbf{r})$ diverges at $r=a$. Outside this interval the cotangent function makes the potential oscillates between regions with negative and positive divergences with $V_0$ positively defined, which difficult the analysis and we will not consider this situation in the present work.

\begin{figure*}
\begin{minipage}[!b]{0.48\linewidth}
\includegraphics[width=.9\linewidth]{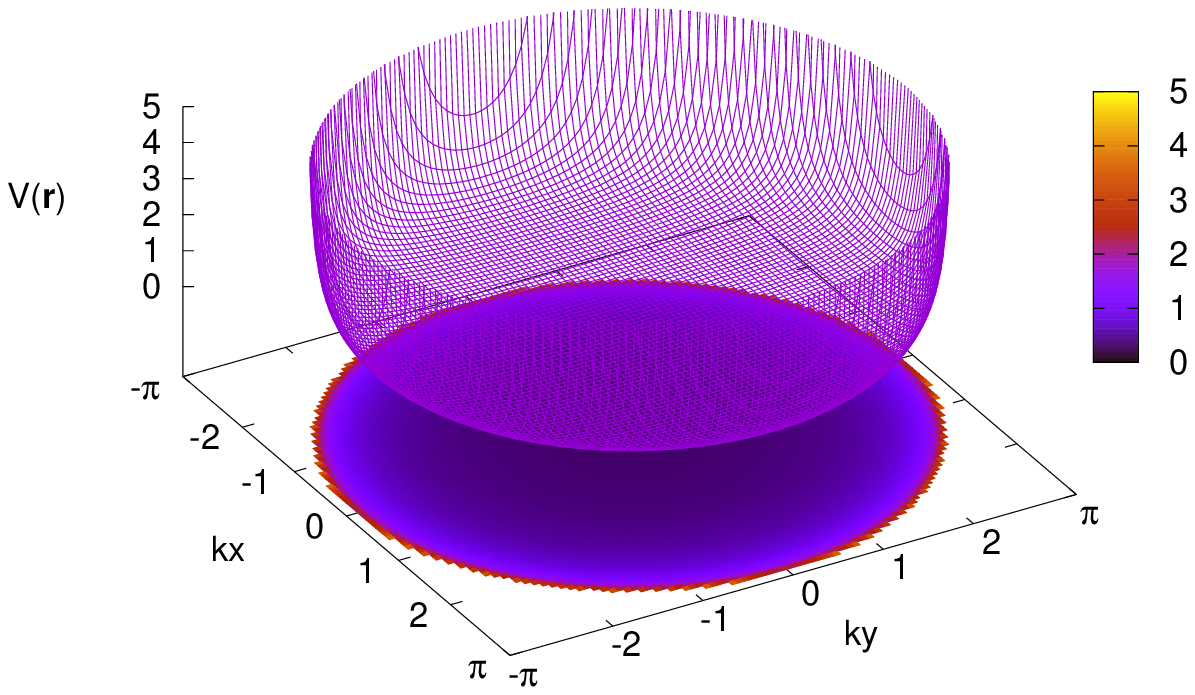}\vspace{1.0cm}
\caption{Surface plot of the mathematical potential in Eq. (\ref{pot3D}).}\label{potential3D}
\end{minipage}\hfill%
\begin{minipage}[!b]{0.48\linewidth}
\centering
\includegraphics[width=.9\linewidth]{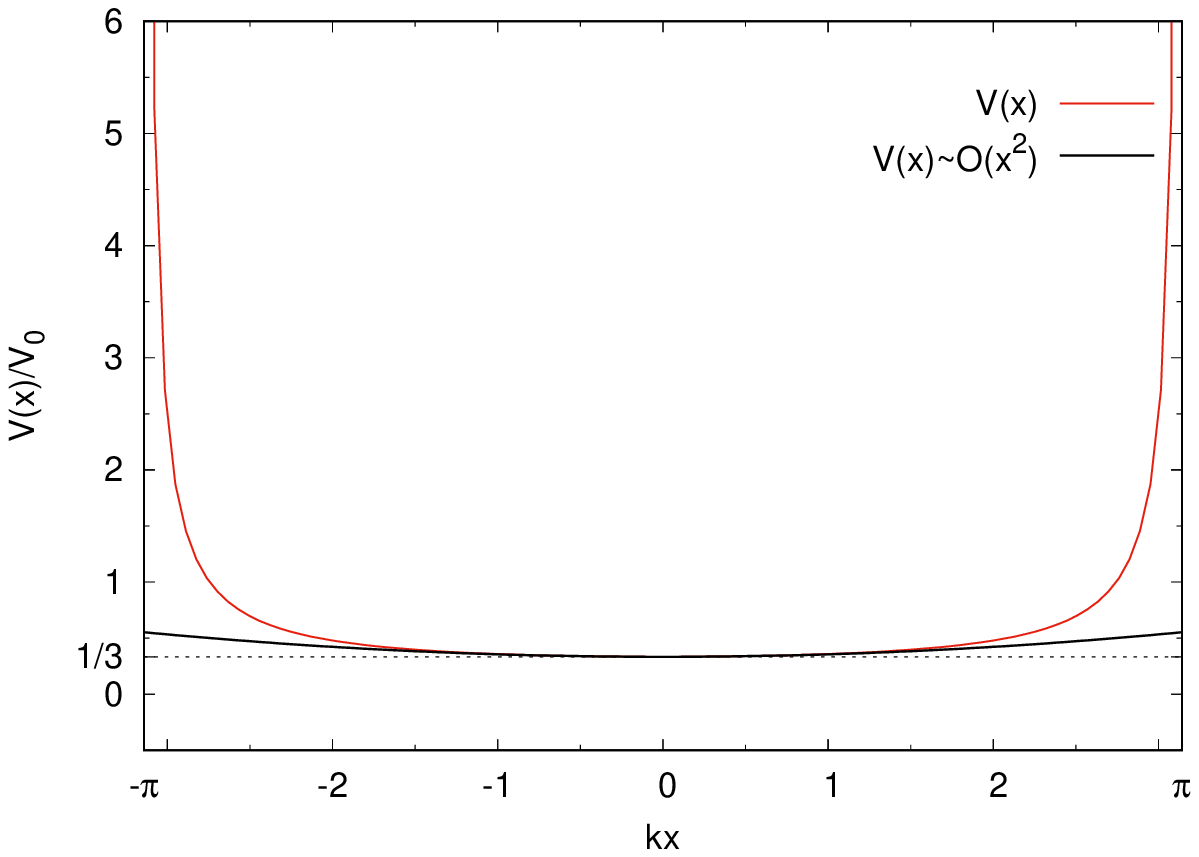}\vspace{1.0cm}
\caption{1d plot of the potential.}\label{potential1d}
\end{minipage}%
\end{figure*}

In figure \ref{potential3D} we display the 3D plot for $z$-coordinate equals zero. If one varies $z$, the diameter of the circle at the bottom of the figure becomes narrower as $|z|$ grows up to a maximum value $z=a$, making $x=0$ and $y=0$ such that $r=a$, which is a point of divergence at the center of the circle. For pedagogical applications, the $z$-coordinate will be set zero from now on. Note that the surface figure generated is formed basically by symmetrical curves with minimum values.       
      
In particular, we are interested in those curves with minimum values given by

\begin{eqnarray}
\lim\limits_{r\rightarrow0}V(\mathbf{r})=\frac{V_0}{3}\label{limite}
\end{eqnarray}

\noindent i.e., those curves that passes through the point $(0,0,0)$.  Let us take the curve belonging to the plane  $(x,0,0)$, which has the limit (\ref{limite}) above and, consequently, $r=|x|$. However, as we already mentioned, Eq. (\ref{pot3D}) is spherically symmetric, which means that if we substitute $r=|x|$ by $-r=-|x|$, the potential remains the same. As a consequence we can drop the absolute value of $|x|$ and write only $x$ to have a 1d version of this mathematical potential

\begin{eqnarray}
V(x)=V_0\left(\frac{1-kx\cot(kx)}{(kx)^2}\right)\label{pot1D}
\end{eqnarray} 

This 1d potential is plotted in figure \ref{potential1d}. As one can see, in the symmetrical interval $kx\in[-\pi,\pi]$, it takes a minimum value at $V_0/3$ and diverges at $\pm\pi$. This is the potential we will use to solve the 1d Schr\"odinger equation. 
To assess the singularities of this potential we firstly consider the series expansion of the cotangent function around the origin given by

\begin{eqnarray}
 \cot{kx} = \frac{B_0}{kx} + \sum_{n=1}^{\infty}&&\frac{(-1)^n 2^{2n}B_{2n}(kx)^{2n-1}}{(2n)!}B_{2n},\nonumber \\
  &&\hspace{2.5cm}0 < |kx| < \pi
\end{eqnarray}

\noindent where $B_{2n}$ are the Bernoulli numbers: $B_0 = 1$, $B_1 = -\frac{1}{2}$,  $B_2 = \frac{1}{6}$, $B_4 = -\frac{1}{30}$, $B_6 = \frac{1}{42}$, $B_8 = \frac{-1}{30}$, etc. As expected, this function exhibits a singularity at the origin. If one multiplies both sides by $kx$, this series can be rewritten as
 
\begin{equation}
 kx \cot{(kx)} = B_0 + \sum_{n=1}^{\infty}\frac{(-1)^n (2kx)^{2n}}{(2n)!}B_{2n}
\end{equation}

\noindent where the singularity is now removed. If we use the value of $B_0=1$ and divide both sides by $(kx)^2$ with $x\neq0$, we can rewrite this series as

\begin{eqnarray}
 \left(\frac{1- kx \cot{(kx)}}{(kx)^2} \right) = - 4\sum_{n=1}^{\infty}&&\frac{(-1)^n (2kx)^{2n-2}}{(2n)!}B_{2n}, \nonumber \\
 &&\hspace{2cm}|kx|<\pi \label{potencial_1}
\end{eqnarray}

This series converges for all the values in the interval $kx\in]-\pi,\pi[$, as one can check taking the $x\rightarrow0$ limit in Eq. (\ref{potencial_1}). Thus, the 1d potential can be expressed as

\begin{equation}
\label{potencial_1d_2}
V(x) =
  \begin{cases}
    -4V_0 \sum\limits_{n=1}^{\infty}\frac{(-1)^n (2kx)^{2n-2}}{(2n)!}B_{2n}  & \quad \text{if } |kx|<\pi\\
    \quad \infty  & \quad \text{if } |kx|=\pi
  \end{cases}
\end{equation}

The first terms of the infinite series are given by

\begin{eqnarray}
\sum_{n=1}^{\infty}\frac{(-1)^n (2kx)^{2n-2}}{(2n)!}B_{2n} &&= -\frac{B_2}{2} + \frac{B_4}{6}(kx)^2 - \nonumber \\
&& \frac{B_6}{45}(kx)^4 + \frac{B_8}{630}(kx)^6 + \cdots \label{serie}
\end{eqnarray}

Notice that the potential will be given in even exponents of the series, which favors the use of the harmonic oscillator Hamiltonian as the unperturbed Hamiltonian when perturbation theory is employed. The potential plot in the $O(x^2)$ approximation is also displayed in figure \ref{potential1d}. We will see the employment of approximated methods in the next sections.

\section{1d Sch\"odinger equation}\label{1d_schrod}

The 1d Schr\"odinger equation is given by 

\begin{equation}
\label{EDO_1}
 -\frac{\hbar^2}{2m} \frac{d^2\psi(x)}{dx^2} + V(x)\psi(x) = E \psi(x)
\end{equation} 

For this particular potential the 1d Schr\"odinger equation can be solved analytically in a very special situation. To see that, let us consider the potential given by Eq. (\ref{pot1D}) and define a new variable $\xi=kx$ such that Eq. (\ref{EDO_1}) can be rewritten as

\begin{equation}
\label{EDO_2}
 \frac{d^2\phi(\xi)}{d\xi^2} - 2 \upsilon_0\left( \frac{1-\xi\cot(\xi)}{\xi^2}\right)\phi(\xi) = - 2 \varepsilon \phi(\xi)
\end{equation}

\noindent where $\phi(\xi)\equiv\psi(\xi/k)$, $\upsilon_0=\frac{mV_0}{\hbar^2 k^2}$, and $\varepsilon = \frac{mE}{\hbar^2 k^2}$. We want to solve this equation for $\xi$ with the boundary condition $\phi(\pm\pi)=0$, where the potential diverges. Note that $\upsilon_0$ is dimensionless. $V_0$ can assume any positive value in dimensions of $\hbar^2 k^2/m$ provided that $E$ is greater than $V_0/3$. For simplicity we consider this positive value equals 1 from now on. 

It is straightforward and important to notice that,

\begin{equation}
 \frac{1-\xi\cot(\xi)}{\xi^2} = -\frac{1}{\sin(\xi)}\frac{d}{d\xi}\left(\frac{\sin(\xi)}{\xi}\right),
\end{equation}

then we plug the above identity in Eq. (\ref{EDO_2}) to obtain 

\begin{equation}
\label{EDO_bessel2}
 \frac{d^2\phi(\xi)}{d\xi ^2} + 2\upsilon_0 \frac{\phi(\xi)}{\sin(\xi)}\dfrac{d}{d\xi}\left( \frac{\sin(\xi)}{\xi}\right) +  2\varepsilon \phi(\xi) =0.
\end{equation}

As we can note, if we choose a Spherical Bessel Function $j_0(\xi) = \sin(\xi)/\xi$ we obtain,

\begin{equation}
\label{EDO_bessel3}
 \frac{d^2 }{d\xi ^2} j_0(\xi) + \frac{2\upsilon_0}{\xi} \dfrac{d }{d\xi} j_0(\xi)+  2\varepsilon  j_0(\xi) = 0.
\end{equation}

In this sense we must impose $\upsilon_0 =1$ and $2\varepsilon =1$ to obtain a common solution for Spherical Bessel Function  of the first kind of the zeroth-order. Consequently we have

\begin{equation}
\label{V_0_e_E}
V_0 = \frac{\hbar^2 k^2}{m} \hspace*{10pt} \mbox{and}  \hspace*{10pt} E= \frac{\hbar^2 k^2}{2m}\,, 
\end{equation}

valid for a wave function $\phi(\xi) = A\sin(\xi)/\xi$. We shall demonstrate \textit{a posteriori} that this is actually the ground state wave function with eigenvalue $E = 0.5 \hbar^2 k^2/m$ using a numerical approach to solve the 1d Schr\"odinger equation (\ref{EDO_1}).

Now it is a good exercise to employ some approximate approach to have an idea about the energy levels for the bound states without too much effort. The  Wentzel-Kramers-Brillouin (WKB) approximation is perfect for such a situation, because the bounded particle, in this case, has a sufficiently high momentum around the center of the potential as well as we can expect that its wave function varies rapidly with position, much more rapidly than the potential \cite{Weinberg}, as is the case when one considers the particle in a box.
  
To this end we can approximate the potential to $O(x^2)$ with the aid of Eq.'s (\ref{potencial_1d_2}) and (\ref{serie}), as plotted in figure \ref{potential1d}. The boundary condition for the potential well in Eq. (\ref{EDO_1}) with WKB method is such that $V(\pm\pi) \rightarrow \infty$ implies $\phi(\pm\pi)=0$. In the pedagogical exercise of one-dimensional square well potential with perfectly rigid walls in a symmetrical interval $-a\leq x \leq a$ where $V(x)=0$ and $V(\pm a)=\infty$, the WKB method gives us an exact solution of energy levels \cite{Mahon,Griffths,Schiff}. Clearly the potential (\ref{potencial_1d_2}) has rigid walls in $-\pi\leq kx \leq \pi$ in $O(x^2)$(as can be seen in figure \ref{potential1d}), so that we can use WKB method to obtain approximated energy levels. Thus, in the classical region of the potential $E>V(x)$, the WKB method gives the following solution 

\begin{equation}
\label{aproximacao_WKB_1}
 \psi(x) \propto \frac{C}{\sqrt{p(x)}}\, e^{\pm i\sigma(x)} 
\end{equation}

where $\sigma(x) = \frac{1}{\hbar} \int p(x) dx$, and $p(x)$ is the momentum given by

\begin{equation}
\label{momento_1}
 p(x) =  \sqrt{2m[E- V(x)]}. 
\end{equation}

The linear combination of solutions (\ref{aproximacao_WKB_1}) is also a solution and

\begin{equation}
 \psi(x) \approx \frac{1}{\sqrt{p(x)}}\,\left[C_1 e^{i\sigma(x)} + C_2 e^{-i\sigma(x)} \right].\nonumber
\end{equation}

Due to the symmetry of the well $-\pi\leq kx \leq \pi$, it admits even wave functions 

\begin{equation}
\label{solucao_par_1}
 \psi_{even}(x) \approx \frac{A}{\sqrt{p(x)}}\, \cos[\sigma(kx)],
\end{equation}

and odd wave functions

\begin{equation}
\label{solucao_impar_1}
 \phi_{odd}(x) \approx \frac{B}{\sqrt{p(x)}}\, \sin[\sigma(kx)].
\end{equation}

As already mentioned, the potential (\ref{potencial_1d_2}) has rigid walls in $kx = \pm \pi$ and the wave function must be zero at these points. For even wave function, $\phi_{even}(\pi)=0$, we have

\begin{equation}
 \sigma(\pi)=\frac{\pi}{2}, \frac{3\pi}{2}, \frac{5\pi}{2},\cdots
\end{equation}

while for the odd wave function, $\phi_{odd}(\pi)=0$, we have

\begin{equation}
 \sigma(\pi)=\frac{2\pi}{2}, \frac{4\pi}{2}, \frac{6\pi}{2},\cdots
\end{equation}

so we assign

\begin{equation}
 \sigma(\pi)=\frac{n \pi}{2} \hspace*{6pt} \mbox{and} \hspace*{6pt}  \sigma(-\pi)=-\frac{n \pi}{2} \hspace*{6pt}\mbox{where} \,\, n=1,2,3,\cdots \nonumber
\end{equation}

It follows that 

\begin{equation}
 \sigma(\pi) - \sigma(-\pi) = \frac{1}{\hbar} \int_{-\pi/k}^{\pi/k} p(x) \, dx, \nonumber
\end{equation}

then it is straightforward to obtain

\begin{equation}
\label{momento_2}
 \frac{1}{\hbar} \int_{-\pi/k}^{\pi/k} p(x) dx = n\pi \hspace*{1cm} \mbox{with} \,\, n=1,2,3,\cdots  
\end{equation}

Now one can write explicitly the potential (\ref{potencial_1d_2}) to $O(x^2)$

\begin{equation}
\label{energia_potencial_3}
V(x) \approx   V_0 \left( \frac{1}{3} +\frac{1}{45} (kx)^2 \right),
\end{equation}

\noindent where in the symmetrical interval $[-\pi,\pi]$ we have $O(1) \gg O(x^2) \gg O(x^4)$, such that the momentum (\ref{momento_1}) can be written as

\begin{equation}
 p(x)   \approx \sqrt{2m\left(E- \frac{1}{3}\frac{\hbar^2 k^2}{m}\right)}\left(1-\frac{(\frac{\hbar^2 k^2}{m})(kx)^2 }{90\left(E- \frac{1}{3}\frac{\hbar^2 k^2}{m}\right) } \right). \nonumber
\end{equation}

and the integral (\ref{momento_2}) can now be calculated

\begin{equation}
 \frac{1}{\hbar}\sqrt{2m\left(E- \frac{1}{3}\frac{\hbar^2 k^2}{m}\right)}\left[\frac{2\pi}{k}-\frac{(\frac{\hbar^2 k^2}{m})\frac{\pi^3}{k} }{135\left(E- \frac{1}{3}\frac{\hbar^2 k^2}{m}\right) } \right] = n\pi, \nonumber
\end{equation}

it follows that

\begin{equation}
\label{energias_WKB_2}
 E_n = \left[\frac{1}{8}\left( \frac{4\pi^2}{135}+\frac{n^2}{2} + \sqrt{\frac{4\pi^2}{135}n^2+\frac{n^4}{4}   }\right) + \frac{1}{3} \right]\frac{\hbar^2 k^2}{m}.
\end{equation}

\noindent where we put the label $n$ in $E_n$ to indicate that the energy depends on $n$. Note that for $n\rightarrow\infty$, $E_n\sim n^2$ as we can expect for a particle in a box. This approximation gives a good idea about the energy levels although the corrections do not differ significantly from the energy levels of the particle in a box. For example, consider $n=1$ in Eq. (\ref{energias_WKB_2}). One obtains $E_1\approx 0.5244 \hbar^2 k^2/m$, while for a particle in a box with an energy shift of $V_0/3$ and a length $a$ gives $0.4583\hbar^2 k^2/m$. For $n=8$ one obtains $E_8=8.406 \hbar^2 k^2/m$ and $8.333 \hbar^2 k^2/m$ for a particle in a box. 

It is straightforward and important to notice that the ground state energy, for $n=1$, is approximately $E = 0.5 \frac{\hbar^2 k^2}{m}$ as seen in Eq. (\ref{V_0_e_E}). The WKB approach indicates the approximate values of energy of a quantum system, but as we have seen it is necessary to solve integrals like equation (\ref{momento_2}), with $p(x) =  \sqrt{2m[E- V(x)]}$, where there are some difficulties to calculate it when the potential is a function expanded in a power series. 

One must employ the WKB approach calculating the turning points correctly for each energy, which is a tough task once these points depend on the energy, and the condition (\ref{momento_2}) does not apply anymore. However, this will be left to the section results only for comparison reasons. For now, the resulting equation (\ref{energias_WKB_2}) gives us a good idea about the energy levels for our problem. In the next section, we shall see the Stationary Perturbation Theory where we obtain approximated values of energy and the wave functions for some specific cases.

\section{Stationary Perturbative Theory}\label{SPT}

We have already encountered from equations (\ref{potencial_1d_2}) and (\ref{serie}) that the expansion of potential contains terms like $x^2$, $x^4$, and so on. Thus, it is useful to make an approach with the Stationary Perturbation Theory to study the effect of the perturbed potential with $x^4$ and $x^6$ on the energy levels into the one-dimensional harmonic oscillator Hamiltonian\cite{Griffths,Schiff,Cohen,Sakurai}. As we shall see, the Stationary Perturbation Theory is also a good technique to compute the first levels of energy and the respective wave functions\cite{feynman}. However, the calculations can be very tough as soon as we go to higher levels of energy. Needless to say that the respective wave functions are also cumbersome to compute. This is due to the fact that the potential (\ref{potencial_1d_2}) is an infinite series and we can only probe some terms of the series. To circumvent this inconvenience one can resort to numerical techniques as we will see in the next section. Nevertheless, it remains important to understand the basic physics of the approximate solutions even before we explore a more accurate numerical method. Let us write the potential well (\ref{potencial_1d_2}) expanded to sixth order,

\begin{equation}
V(x) \approx V_0 \left( \frac{1}{3} +\frac{1}{45} (kx)^2 + \frac{2}{945}(kx)^4 + \frac{1}{4725}(kx)^6 \right),\nonumber
\end{equation}

as we already set $V_0=\dfrac{\hbar^2 k^2}{m}$, we write the unperturbed Hamiltonian as 

\begin{equation}
{-\frac{\hbar^2}{2m} \frac{d^2\psi}{dx^2}  + \frac{1}{45} \frac{\hbar^2 k^2}{m}  (kx)^2 \psi} ={H_0 \psi}
\end{equation}

where $H_0$ is the unperturbed Hamiltonian, which is the Hamiltonian for an 1d Harmonic Oscillator with $\frac{1}{45} \frac{\hbar^2k^4}{m} =\frac{1}{2}m\Omega^2$.

For the perturbing potential, $W$, we write

\begin{equation}
W(x)\psi =  \left(\frac{1}{3} +\frac{2}{945} (kx)^4 +\frac{1}{4725} (kx)^6\right) \frac{\hbar^2 k^2}{m} \psi.
\end{equation}

Using the transformation $\xi=kx$ we now write the 1d Sch\"odinger equation as

\begin{equation}
- \frac{d^2\phi(\xi)}{d\xi ^2}  + \frac{1}{2}\omega^2\xi^2 \phi(\xi) +
\left(\frac{1}{3} + \frac{2\xi^4}{945}  +\frac{\xi^6}{4725}\right)  \phi(\xi) = \varepsilon \phi(\xi)
\end{equation}
where we have $\omega=\frac{m\Omega}{\hbar k^2}\equiv\sqrt{\frac{2}{45}}$ and $\varepsilon = \frac{mE}{\hbar^2 k^2}$ dimensionless constants.

Now, it is convenient to define the creation and annihilation operators respectively \cite{Cohen,feynman},

\begin{equation}
\label{operadores_1}
a^{\dag} = \sqrt{\frac{\omega}{2}}\left(\xi-\frac{ip}{\omega}\right), \hspace*{25pt} a = \sqrt{\frac{\omega}{2}}\left(\xi + \frac{ip}{\omega}\right),
\end{equation}

\noindent where we define the operator $p=-i\frac{d}{d\xi}$. These operators obey the rules below
 
\begin{equation}
\label{operadores_2}
a^{\dag} a = N,\hspace*{15pt} [a,a^{\dag} ] = 1, \hspace*{15pt} [N,a^{\dag} ] = a^{\dag},\hspace*{15pt} [N,a] = -a,
\end{equation}

\noindent also we have

\begin{equation}
\label{operadores_3}
a\,| n \rangle = \sqrt{n} \,| n -1 \rangle \hspace*{15pt} \mbox{and} \hspace*{15pt} a^{\dag}\, | n \rangle = \sqrt{n+1} \,| n +1 \rangle .
\end{equation}

The unperturbed Hamiltonian of the one-dimensional harmonic oscillator yields energy values 

\begin{equation}
\label{energia_nao_perturbada}
 \varepsilon_{n}^{(0)} = \left(n +\frac{1}{2}\right)\omega,
\end{equation}

\noindent with $\varepsilon_{n}^{(0)}\equiv\frac{mE^{0}}{\hbar^2k^2}$, where the Stationary Perturbation Theory will add corrections to (\ref{energia_nao_perturbada}).

It is straightforward to obtain the coordinate operators using the equations (\ref{operadores_1})

\begin{equation}
\xi = \sqrt{\frac{1}{2\omega}}\left(a+ a^{\dag} \right),
\end{equation}

\noindent and with the aid of commutation relations (\ref{operadores_2}) we can get

\begin{eqnarray}
\label{x_quarta}
 \xi^4  = \left(\frac{1}{2\omega}\right)^2 & &[a^4+ a^{\dag\, 4} + (4N+6) a^2 + (4N-2)a^{\dag\, 2}\cr & & +(6N^2+ 6N+3)]
\end{eqnarray}

\noindent and
 
\begin{eqnarray}
\label{x_sexta}
 \xi^6  &=& \left(\frac{1}{2\omega}\right)^3  \big[a^6 + a^{\dag\, 6} +(6N+15)a^4 + (6N-9)a^{\dag\, 4}\cr  & & +(15N^2 +45N +45)a^2 
 + (15N^2 -15N +15)a^{\dag\, 2}\cr 
 & & +(20N^3+30N^2+40N+15) \big].
\end{eqnarray}

According to Stationary Perturbation Theory, the first-order correction to the energy is simply equals the mean value of perturbation term W in the unperturbed state $| n \rangle$,

\begin{equation}
\label{perturbing_1}
\varepsilon^{(1)}_{n} = \langle n|\, W \, | n \rangle.
\end{equation}

\noindent where we set the perturbing term as

\begin{equation}
\label{perturbing_2}
 W= \frac{1}{3} + \frac{2\xi^4}{945} + \frac{\xi^6}{4725}
\end{equation}

We can see that the only non-vanishing terms in the first-order correction to the energy in Eq. (\ref{perturbing_1})(when we use the equations (\ref{x_quarta}) and (\ref{x_sexta})) are $6N^2+ 6N+3$ and $20N^3+30N^2+40N+15$, such as

\begin{eqnarray}
\varepsilon^{(1)}_{n} &=& \frac{1}{3} + \frac{2}{945} \left(\frac{1}{2\omega}\right)^2(6n^2+6n +3)\cr
& & + \frac{1}{4725} \left(\frac{1}{2\omega}\right)^3 (20n^3+30n^2+40n+15) . \nonumber
\end{eqnarray}

Replacing $\omega^2 =\dfrac{2}{45}$ into the above equation we can write

\begin{eqnarray}
\label{1a_correcao_energia}
\varepsilon^{(1)}_{n}  &=&  \frac{1}{3} + \frac{1}{84} (6n^2+6n +3) \cr
 & & + \sqrt{\frac{45}{2}}\,\left(\frac{1}{1680}\right)(20n^3 
 +30n^2+40n+15).\cr & &
\end{eqnarray}

As we mentioned elsewhere, the calculations for higher levels in Stationary Perturbation Theory is a tough task with a great quantity of integral calculations. There is no reason to go beyond once several accurate numerical techniques are available to solve the Sch\"odinger equation. As a consequence, we are limited to the first levels of energy where the first terms in the potential expansion are important. For this reason we will disregard corrections due to the sixth-order term $\xi^6$, we will simplify the calculations to second-order of correction to the energy using just only the term $\xi^4$ of (\ref{perturbing_2}) for both energy calculations and the respective wave functions, keeping this contribution only in the first-order correction as shown in Eq. (\ref{1a_correcao_energia}). We therefore have

\begin{equation}
\varepsilon^{(2)}_n = \sum_{n\neq n'} \frac{|\,\langle n|\, W_{\xi^4} \, | n' \rangle\,|^2}{\varepsilon_{n'}^{(0)} - \varepsilon_{n}^{(0)}}.
\end{equation}

The terms of $W_{\xi^4} = \frac{2\xi^4}{945}$ that contribute to second-order are $\left(\dfrac{1}{2\omega}\right)^2[a^4+ a^{\dag\, 4} + (4N+ 6) a^2   +  (4N-2) a^{\dag\, 2}] $, so that it follows

 \begin{eqnarray}
 \langle n-2| W_{\xi^4}| n \rangle &=&  \frac{1}{84} \left[ (4n-2) \sqrt{n(n-1)} \right] 
 \cr
 \langle n+2| W_{\xi^4}| n \rangle &=& \frac{1}{84}  \left[ (4n+6) \sqrt{(n+1)(n+2)} \right] \cr
 \langle n-4| W_{\xi^4}| n \rangle &=& \frac{1}{84} \left[ \sqrt{n(n-1)(n-2)(n-3)}  \right]
 \cr
 \langle n+4| W_{\xi^4}| n \rangle &=& \frac{1}{84}  \left[ \sqrt{(n+1)(n+2)(n+3)(n+4)}, \right]\cr & &
\end{eqnarray}

The values of the matrix elements $\langle n|\, W_{\xi^4} \, | n' \rangle$ necessary to compute the energy corrections as well as the respective wave functions are listed in Appendix \ref{coeficientes}. In what follows, when we calculate the second-order correction for the ground state energy we find

\begin{eqnarray}
\label{E_fundamental_2}
 \varepsilon^{(2)}_0 &=& \frac{| \langle 2| W_{\xi^4}| 0 \rangle|^2 }{ \varepsilon^{(0)}_0 -  \varepsilon^{(0)} _2} +  \frac{| \langle 4| W_{\xi^4}| 0 \rangle| ^2}{ \varepsilon^{(0)}_0 -  \varepsilon^{(0)} _4}  \cr &=& \frac{\frac{2}{196} }{(-2)\sqrt{\frac{2}{45}}} +   \frac{\frac{6}{1764} }{(-4)\sqrt{\frac{2}{45}}} \cr &=& -0.028234621.
\end{eqnarray}

We can continue with this method and obtain some values for second-order correction values such as 

\begin{eqnarray}
\label{E_correcao_2a_ordem}
 \varepsilon^{(2)}_1 &=& -0.2218,\hspace*{10pt}
 \varepsilon^{(2)}_2 = -0.8269 \cr
 \varepsilon^{(2)}_3 &=& -2.1176,\hspace*{10pt}  \varepsilon^{(2)}_4 = -4.3683 \cr
  \varepsilon^{(2)}_5 &=& -7.8533,\hspace*{10pt}  \varepsilon^{(2)}_6 = -12.8468 \cr 
  \varepsilon^{(2)}_7 &=& -19.6231
\end{eqnarray}

Therefore we can compute the energy values using (\ref{energia_nao_perturbada}), (\ref{1a_correcao_energia}), (\ref{E_fundamental_2}), and (\ref{E_correcao_2a_ordem}) where we have

\begin{equation}
 \varepsilon_n =  \varepsilon_{n}^{(0)} + \varepsilon_{n}^{(1)} + \varepsilon_{n}^{(2)}
\end{equation}

\noindent with $\varepsilon^{(i)} = \frac{mE^{(i)}}{\hbar^2 k^2},\,\,i=0,1,2$. 

Now we move on to obtain the approximate wave function. The first-order correction of the state vector is a linear superposition of all the unperturbed states

\begin{equation}
\label{auto_funcao_pertubacao}
 \psi_n^{(1)}(x) = \psi_n^{(0)}(x)+ \sum_{n\neq k}\frac{\langle \psi_k^{(0)} |W| \psi_n^{(0)} \rangle}{E_n^{(0)} - E_k^{(0)}}\psi_k^{(0)}.
\end{equation}

The eigenfunctions of the unperturbed  one-dimensional harmonic oscillator are 

\begin{equation}
\label{auto_funcao_nao_pertubacao}
 \psi_n^{(0)}(u) = \left(\frac{m\omega}{\pi\hbar} \right)^{1/4}\frac{1}{\sqrt{2^n\, n!}}H_n(u)e^{-u^2/2},
\end{equation}

\noindent where $u = \sqrt{\frac{m\omega}{\hbar}}\, x\equiv\left(\frac{2}{45}\right)^{1/4}\xi$ and $H_n(u)$ are the Hermite polynomials,
\begin{equation}
 \begin{cases}
  H_0 = 1\cr
  H_1 = 2u \cr
  H_2 = 4u^2 -2 \cr
  H_3 = 8u^3 - 12u \cr
  H_4 = 16u^4 - 48u^2 +12 \cr
  H_5 = 32u^5 -160u^3 +120u.
 \end{cases}
\end{equation}

Thus, we can calculate the ground state wave function using the equation (\ref{auto_funcao_pertubacao}), it follows that

\begin{eqnarray}
\psi_0^{(1)} &=& \psi_0^{(0)} +  \frac{\langle \psi_2^{(0)} |W_{\xi^4}| \psi_0^{(0)} \rangle}{E_0^{(0)} - E_2^{(0)}}\psi_2^{(0)} +  \frac{\langle \psi_4^{(0)} |W_{\xi^4}| \psi_0^{(0)} \rangle}{E_0^{(0)} - E_4^{(0)}}\psi_4^{(0)}\cr\cr
&=& \psi_0^{(0)} +  \frac{\langle 2 |W_{\xi^4}| 0 \rangle}{E_0^{(0)} - E_2^{(0)}}\psi_2^{(0)} +  \frac{\langle 4 |W_{\xi^4}| 0 \rangle}{E_0^{(0)} - E_4^{(0)}}\psi_4^{(0)}\cr\cr
&=& \psi_0^{(0)} +  \frac{\frac{\sqrt{2}}{14}}{(-2)\sqrt{\frac{2}{45}}}\psi_2^{(0)} +  \frac{\frac{\sqrt{6}}{42}}{(-4)\sqrt{\frac{2}{45}}}\psi_4^{(0)}\cr\cr
&=& \psi_0^{(0)} -0.239578711\, \psi_2^{(0)} - 0.069160416\, \psi_4^{(0)}, \nonumber
\end{eqnarray}

\noindent we can now put the values of unperturbed functions of (\ref{auto_funcao_nao_pertubacao}) into above result and obtain

\begin{eqnarray}
\label{estado_fundamental_2}
 \psi_0^{(1)} &=& \left(\frac{1}{\pi}\sqrt{\frac{2}{45}} \right)^{1/4}e^{-u^2/2}[1 - 0.084703865(4u^2 - 2)\cr & & - 0.003529328(16u^4 - 48u^2 +12)].
\end{eqnarray}

In the same way we can obtain the first excited wave function as fallows

\begin{eqnarray}
 \psi_1^{(1)} &=& \left(\frac{1}{\pi}\sqrt{\frac{2}{45}} \right)^{1/4}e^{-u^2/2}[1.414213562u\cr
& & - 0.099824463(8u^3 - 12u)\cr
& & - 0.002495612(32u^5 - 160u^3 +120u)].
\end{eqnarray}

\noindent where these two wave functions are not normalized yet. As we already mentioned, the Perturbative Method is unproductive to obtain accurate results because it is necessary many calculation that converge slowly. For this problem with potential well (\ref{potencial_1d_2}), we will employ an efficient numerical method which can be implemented in any computer language.

\section{Numerical Method}\label{Num_Met}

For simplicity, we will write the 1D Schr\"odinger Equation exactly like we did in Eq. (\ref{EDO_2}), where $\upsilon_0\equiv\frac{mV_0}{\hslash^2k^2}$ and $\varepsilon\equiv\frac{mE}{\hslash^2k^2}$. Unlike what we did in the last sections, we will use from now on $x$ as dimensionless variable instead of $\xi$ for convenience.

\begin{equation}
 \label{eq:Schroedinger}
 -\frac{1}{2}\frac{d^2}{dx^2}\psi +\upsilon(x)\psi = \varepsilon\psi
\end{equation}
The Numerov Method~\cite{Hairer} can always be applied if 
the differential equation is in the format
\begin{equation}
\frac{d^2\psi}{dx^2} = -g(x)\psi +s(x)
\label{eq:numerovDiff}
\end{equation}
\noindent Then we want to solve the equation (\ref{eq:Schroedinger}) when the potential $V(x)$ is given by
\begin{equation}
 \label{eq:potencial}
 \upsilon(x) = \upsilon_0\left(\frac{1-x\cot(x)}{x^2}\right)
\end{equation}
\noindent we will rewrite it as 
\begin{equation}
 \label{eq:SchroNumerov}
 \frac{d^2\psi}{dx^2}= -2\left[ \varepsilon - \upsilon(x) \right]\psi
\end{equation}
\noindent and if we make $g(x)=2\left[ \varepsilon - \upsilon(x) \right]$ then
\begin{equation}
 \label{eq:NumerovFormat}
 \frac{d^2\psi}{dx^2}=\psi''= -g(x)\psi
\end{equation}

\noindent in this case the function $s(x)$ appearing in Eq. (\ref{eq:numerovDiff}) is zero. With this in mind, ${x_k=-\pi+hk}$, and ${k=0,1,2,\;\cdots,\;N}$. Also assume that at $x_k$ we have ${\psi_k = \psi(x_k)}$. Since Eq. (\ref{eq:NumerovFormat}) is an homogeneous linear differential equation. Therefore, if $\varphi_1(x)$ is solution to the equation, then is as well $\varphi_2(x) = A\varphi_1(x) $, where $A$ is some real constant. 

\begin{figure}[h!]
 \label{fig:intervaloX}
 \caption{Assume that the interval is discretized $x \in [-\pi,\pi]$ in $N$ equal parts $h = x_{k+1}-x_k$ or equivalently ${h = 2\pi/N}$. \\}
\includegraphics[scale = 0.32]{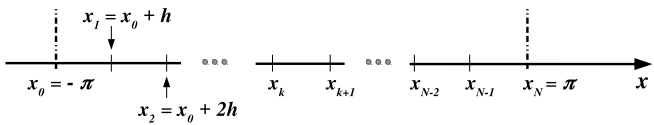}
\end{figure}

It is worth to mention the special case  when $A = -1$ then ${\varphi_2(x) = -\varphi_1(x) }$. This fact will be important because in the Numerov Method we need two consecutive values for the solution, $\psi_k$ and $\psi_{k+1}$. So let us assume we start with $\Psi_0 = 0$ then we can guess $\Psi_1 = \pm \epsilon $. 

From the Boundary Conditions, at $x_0 = -\pi$ we must have $\Psi_0 = 0$, in other words, this point $(-\pi, 0)$ has to be a value for the solution. Guessing a positive or negative value for $\Psi_1$ will only drive us to  $+\Psi(x)$ or $-\Psi(x)$ then it doesn't matter if we guess a positive or negative variation value in the neighborhood of  $\psi_0$ , since we will get either $\Psi$ or $-\Psi$ but both are solutions. Assume that the amplitudes of some solution $\psi$ when $x$ is discretized in the interval $[-\pi,\pi]$  be given by  $\psi=\{\psi_0, \psi_1, \psi_2, \cdots, \psi_N\}$. Accordingly,  ${\Psi =A\psi=\{A\psi_0, A\psi_1, A\psi_2, \cdots, A\psi_N\}}$ constitute also a set of amplitudes which are also amplitudes for another solution of the Schr\"odinger Equation~(\ref{eq:Schroedinger}).

We are free to pick any value for constant $A$, we can always adjust it such that $A\psi_1=\epsilon$, for instance, $\epsilon = 0.0001$, or in other words: $A = 10^{-4}/\psi_1$. That is reasonable-since $\Psi_0 = 0$, and because the solution is assumed to be a smooth function-then, we will not expect huge variations from one value to the ones in some neighborhood. If we can calculate the next value $\Psi_2$ based on $\Psi_0$ and $\Psi_1$, we can repeat the process and find all other values of $\Psi_k$. We can use the Numerov's  Method to find ${\Psi_{k+1}=f(\Psi_k,\Psi_{k-1})}$, which is derived in the Appendix \ref{numerov_algorithm}. This method has order 4 of convergence~\cite{Hairer}.

The selection goes as following, for those values of $E$, starting with $\Psi_0=0 $ and $\Psi_1= \epsilon$ which finish with $\Psi_N \approx 0 $ will constitute a set of valid values for the solution of the Schr\"odinger equation, in this way complying with the two Boundary Conditions in $\pm\pi$ or $\Psi(\pi) = \Psi(-\pi) =0$.
 
\section{Results}\label{Res}

Unlike what was done in section \ref{1d_schrod} we must implement the full WKB method to calculate the energies and the respective wave functions\cite{Mahon}. There we approximated the potential to $O(x^2)$ and derived an expression for $E_n\sim n^2$. As a consequence, the energy levels started from the value of $n=1$. This occurred due to the considerations which led to the expression (\ref{momento_2}). To obtain the energies and the wave functions in WKB for the potential (\ref{pot1D}) we must find the turning points for each energy. This is an additional difficulty because the turning points define the energy. However, one can circumvent this difficulty by applying the following WKB constraint

{\small
\begin{eqnarray}
\label{WKB_const}
 \frac{1}{\hbar} \int_{x_1}^{x_2} \sqrt{2m[E- V(x)]} dx &=& \left(n+\frac{1}{2}\right)\pi \nonumber \\
  &&\mbox{with} \,\, n=0,1,2,\cdots  
\end{eqnarray} 

\noindent where $x_1\equiv x_1(E)$ and $x_2\equiv x_2(E)$ are the turning points. To obtain the energy and the respective turning points we passed the right-hand side in Eq. (\ref{WKB_const}) to the left-hand side, defining a new function. Now the task is to determine the respective values of the energy which makes this new function equals zero, \textit{i.e.}, for a given level $n$ one must only find roots, remembering that the turning points are also dependent of the energy. We did that in Scilab using Newton-Raphson algorithm to find roots, varying the energy with 0.0001$\hbar^2k^2/m$ for each step, calculating the respective turning points making $E=V(x)$, which in turn are used to calculate the integral in Eq. (\ref{WKB_const}). Thus, using the approximated expression for the energy levels in section \ref{1d_schrod} we could assign an initial value below these values to find the energy which minimizes this new function for a given $n$. The resulting energies can be visualized in the first column in Table \ref{table1}. These are the energy values for the full WKB implementation considering the turning points in each level of energy. We will not describe the computation of the respective wave functions here because it is well-established in the text books\cite{Mahon,Griffths}. 

\begin{figure*}[t!]
     \centering
     \subfloat[]{\includegraphics[width=.4\linewidth]{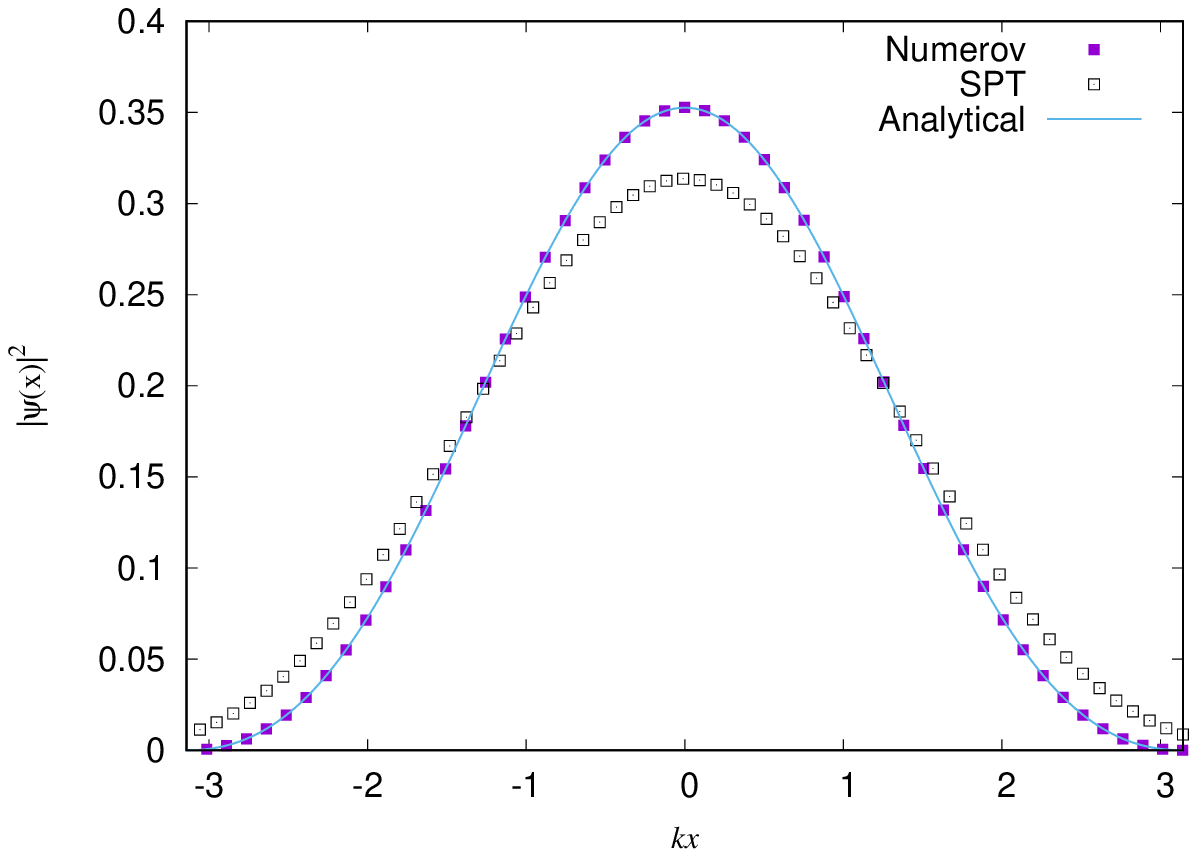}}
     \qquad
     \subfloat[]{\includegraphics[width=.4\linewidth]{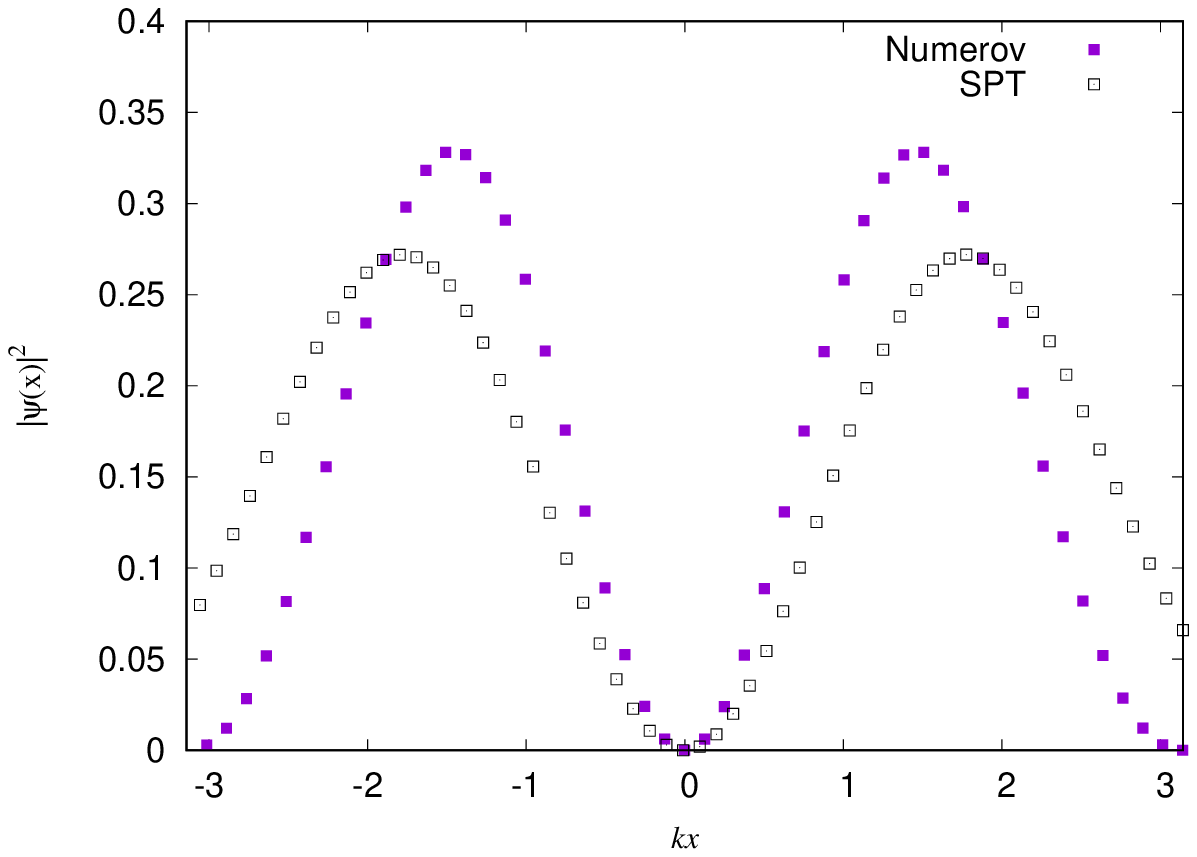}}
     \qquad
    \subfloat[]{\includegraphics[width=.4\linewidth]{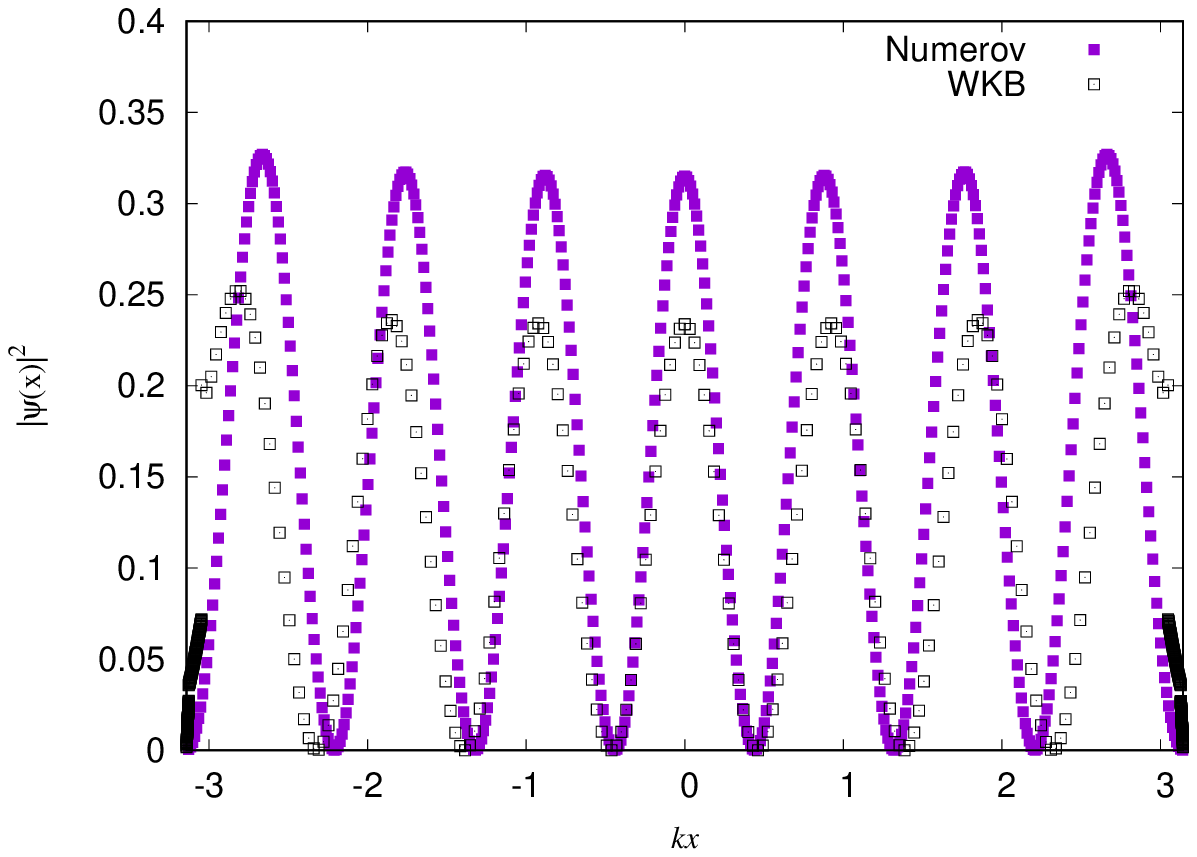}}
     \qquad
     \subfloat[]{\includegraphics[width=.4\linewidth]{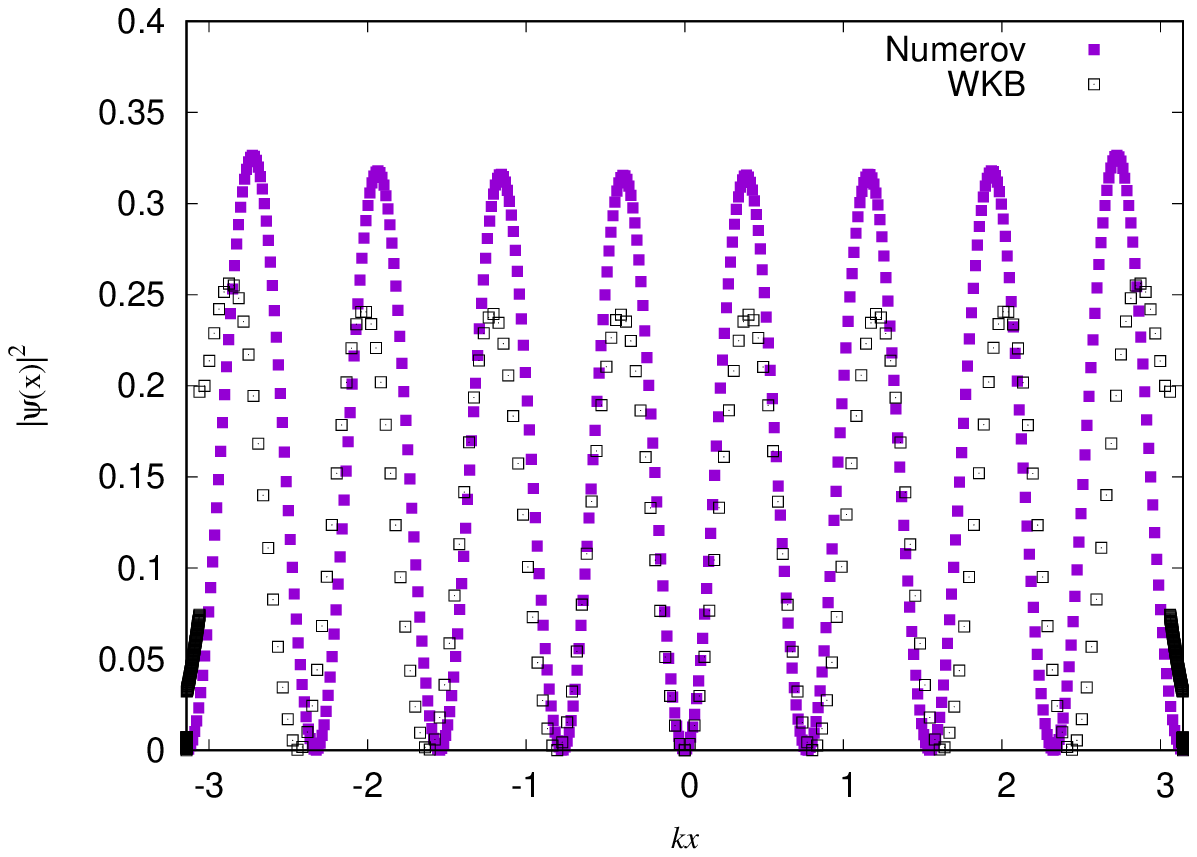}}
      \caption{(a) is the plot for $|\psi|^2$ of the ground state wave function using the analytic expression, and SPT and Numerov methods. (b) is the plot for $|\psi|^2$ of the first excited state for both SPT and Numerov methods. (c) is the plot for $|\psi|^2$ of the sixth excited state for both WKB and Numerov methods. (d)  is the plot for $|\psi|^2$ of the seventh excited state for both WKB and Numerov methods.}\label{figuras}
\end{figure*} 

\begin{table}[t]
\caption{Energies in units of $\frac{\hbar^2k^2}{m}$ calculated using WKB, Stationary Perturbation Theory (SPT) and Numerov's numerical approach for the 1d Schr\"odinger equation with the suggested potential V(x).}
\begin{ruledtabular}
\begin{tabular}{cccc} 
Level $n$ & WKB & SPT & Numerov \\[0.3ex] 
\hline      
0 &0.4579 &0.4886 &0.5000 \\
1 &0.8408  &0.9028  &0.9424 \\
2 &1.4307  & 1.5566 &1.6108 \\
3 &2.2465  & 2.5143 &2.5175 \\
4 &3.2976  & 3.8414 &3.6672 \\
5 &4.5890  & 5.6014 &5.0628 \\
6 &6.1237  & 7.8593 &6.7055 \\
7 &7.9035  & 10.6795 &8.5963 \\
\end{tabular}
\end{ruledtabular}
\label{table1}            
\end{table}  

Now we call attention to the first level of Numerov's algorithm calculation. The value of 0.500 $\hbar^2k^2/m$ is exactly the ground state value obtained analytically in section \ref{1d_schrod}. This method produces good results in solving 1d-Schr\"odinger equation like pointed out by  F. Caruso and V. Oguri\cite{Caruso}. The energy results are summarized in the last column and will serve to compare with other approximated methods. As one can see, the WKB method gives the best results if compared with Stationary Perturbation Theory (SPT) for the higher levels. One of the reasons that the SPT gives some discrepancies is the fact that we only go up to $O(x^4)$ in the potential series to the second-order corrections as we mentioned elsewhere. However, these discrepancies are only perceptive after $n=4$, giving reasonable results for the energy levels bellow that. As expected, the WKB method approximates to the exact values as soon as $n$ becomes big. 

In figure \ref{figuras} we plot $|\psi|^2$ for some wave functions generated by each method. In figure \ref{figuras} (a) we plot the analytic expression for the ground state wave function from section \ref{1d_schrod} with that obtained using SPT and the Numerov methods. As one can see, Numerov is quite accurate while the SPT, although not so accurate, seems to be a reasonable approximation. However, we cannot say the same in figure  \ref{figuras} (b) for the first excited state. The SPT result in comparison with Numerov's solution is quite different due to the increase in the difference of the energies between the methods as can be seen in Table \ref{table1}. Finally, in figures \ref{figuras} (c) and (d) we plot the WKB results for $n=6$ and $n=7$ with the respective Numerov's results. We can see that there is a difference in amplitude and in the phase which can be explained by the difference in energy as can be seen in Table \ref{table1}. 
   
\section{Conclusion}\label{conclu}

We have suggested a mathematical potential as a good exercise in employing different Quantum Mechanical techniques to solve the 1d-Sch\"odinger equation. We applied both WKB and Stationary Perturbation Theory to get the energy states of the 1d-Sch\"odinger equation and compared them with an accurate numerical solution described elsewhere. One can explore other applications of such potential to other physical situations like to solve the 3d-Sch\"odinger equation for certain conditions and even to solids. Other possibilities will be explored in future works.  
 
\appendix

\section{Perturbative Matrix Elements}\label{coeficientes}

Here we give some calculated values of the matrix elements $\langle n|\, W_{\xi^4} \, | n' \rangle$ below. Due to the symmetry $\langle n|\, W_{\xi^4} \, | n' \rangle = \langle n'|\, W_{\xi^4} \, | n\rangle$ all these elements can be used to calculate the transposed ones. 
{\small
\begin{eqnarray}
&& \langle 2| W_{\xi^4}| 0 \rangle = \frac{\sqrt{2}}{14}, \hspace*{3pt}  \langle 4| W_{\xi^4}| 0 \rangle = \frac{\sqrt{6}}{42}, \hspace*{3pt}  \langle 3| W_{\xi^4}| 1 \rangle = \frac{5\sqrt{6}}{42}, \nonumber \\
&&\langle 5| W_{\xi^4}| 1 \rangle = \frac{\sqrt{30}}{42}, \hspace*{3pt} \langle 4| W_{\xi^4}| 2 \rangle = \frac{\sqrt{3}}{3}, \hspace*{3pt} \langle 6| W_{\xi^4}| 2 \rangle = \frac{\sqrt{10}}{14}, \nonumber \\
&& \langle 5| W_{\xi^4}| 3 \rangle = \frac{3\sqrt{20}}{14},  \hspace*{3pt}  \langle 7| W_{\xi^4}| 3 \rangle = \sqrt{\frac{5}{42}},  \hspace*{3pt}  \langle 7| W_{\xi^4}| 5 \rangle = \frac{13}{\sqrt{42}} \nonumber \\
&&\langle 6| W_{\xi^4}| 4 \rangle = \frac{11\sqrt{30}}{42},\hspace*{3pt} \langle 8| W_{\xi^4}| 4 \rangle = \sqrt{\frac{5}{21}}, \hspace*{3pt} \langle 8| W_{\xi^4}| 6 \rangle = \frac{15\sqrt{56}}{42} \nonumber \\
&&\langle 9| W_{\xi^4}| 5 \rangle = \frac{3}{\sqrt{21}},\hspace*{3pt} \langle 9| W_{\xi^4}| 7 \rangle = \frac{17\sqrt{2}}{7}, \hspace*{3pt} \langle 10| W_{\xi^4}| 6 \rangle = \frac{\sqrt{140}}{14} \nonumber \\
&&\langle 11| W_{\xi^4}| 7 \rangle = \frac{\sqrt{55}}{7}. 
\end{eqnarray}}

\section{Numerov's Algorithm}\label{numerov_algorithm}

Let us find out an expression $\Psi_{k+1} = f(x_k, x_{k-1})$. We will start with a Taylor series around $x_k$
 
\begin{multline}
\Psi(x_k+h)=\Psi(x_k)+\frac{\Psi'(x_k)}{1!}h+\frac{\Psi''(x_k)}{2!}h^2+\frac{\Psi'''(x_k)}{3!}h^3+\\ \frac{\Psi''''(x_k)}{4!}h^4+\frac{\Psi'''''(x_k)}{5!}h^5+\mathcal{O}(h^6)
\end{multline}

and following our notation, we can rewrite it as

$$\Psi_{k+1}=\Psi_k+\frac{\Psi'_k}{1!}h+\frac{\Psi''_k}{2!}h^2+\frac{\Psi'''_k}{3!}h^3+\frac{\Psi''''_k}{4!}h^4+\frac{\Psi'''''_k}{5!}h^5+\mathcal{O}(h^6)$$ making the change $h\rightarrow -h$ $$\Psi_{k-1}=\Psi_k-\frac{\Psi'_k}{1!}h+\frac{\Psi''_k}{2!}h^2-\frac{\Psi'''_k}{3!}h^3+\frac{\Psi''''_k}{4!}h^4-\frac{\Psi'''''_k}{5!}h^5+\mathcal{O}(h^6)$$

Now if we add the last two equations

\begin{equation}
\label{eq:FiniteDiff_a}
 \Psi_{k+1}+\Psi_{k-1}=2\Psi_k+\Psi''_k h^2+ \frac{\Psi''''_k}{12}h^4+O(h^6)
\end{equation}

\begin{equation}
\label{eq:FiniteDiff_b}
\Psi''_k = \frac{\Psi_{k+1}-2\Psi_k+\Psi_{k-1}}{h^2}+\mathcal{O}(h^2)
\end{equation}

\noindent from eq.(~\ref{eq:NumerovFormat}) we have  $\Psi''_k= -g_k\Psi_k$, and if we substitute in eq.(~\ref{eq:FiniteDiff_a}) we get

\begin{equation}
\label{eq:FiniteDiff_c}
\Psi_{k+1}-2\Psi_k+\Psi_{k-1}= -g_k \Psi_k  h^2+ \frac{\Psi''''_k}{12}h^4+O(h^6)
\end{equation}

\noindent we can use eq.(~\ref{eq:FiniteDiff_b}) in order to modify $\Psi''''_k$ since 

\begin{align*}
 \Psi''''_k =& (\;\;\;\Psi_k''\;\;)'' \\
 =& (-g_k \Psi_k)'' \\
 =& - \frac{g_{k+1}\Psi_{k+1}-2g_k\Psi_k+g_{k-1}\Psi_{k-1}}{h^2}+\mathcal{O}(h^2).
\end{align*}

\noindent if we substitute in eq.(~\ref{eq:FiniteDiff_c}).

\begin{multline}
\Psi_{k+1}-2\Psi_k+\Psi_{k-1}= -g_k \Psi_k h^2\\ -\frac{\left( g_{k+1}\Psi_{k+1}-2g_k\Psi_k+g_{k-1}\Psi_{k-1}\right)}{12}h^2+O(h^6)
\end{multline}

if we isolate $\Psi_{k+1}$ 

\begin{equation}
\label{eq:NumerovAlgorithm}
 \Psi_{k+1} = \frac{\left( 2-\frac{5h^2}{6}g_k \right)\Psi_k- \left( 1+\frac{h^2}{12}g_{k-1}\right)\Psi_{k-1}}{\left( 1+\frac{h^2}{12}g_{k+1}\right)}
\end{equation}


\begin{thebibliography}{99}
\bibitem{Weinberg} Steven Weinberg, ``Lectures on Quantum Mechanics'', Cambridge University Press, (2013);
\bibitem{Sakurai} J.J. Sakurai, ``Modern Quantum Mechanics'', Addison-Wesley Publishing Company, Inc., 1995;
\bibitem{Mahon} Jos\'e Roberto P. Mahon, ``Mec\^anica Qu\^antica: Desenvolvimento Contempor\^aneo com Aplica\c{c}\~oes'', LTC, Geral edition (2011);
\bibitem{Griffths} D. Griffiths, ``Introduction to Quantum Mechanics'', Cambridge University Press, second edition (2016);
\bibitem{Schiff} L. I. Schiff, ``Quantum Mechanics'', McGraw-Hill Book Company,(1968);
\bibitem{Cohen} C. Cohen-Tannoudji, B. Diu, F. Lalo\"e, {``Quantum Mechanics''}, Volume II, John Wiley \& Sons (1977);
\bibitem{Hairer} Hairer, Ernst; Nørsett, Syvert Paul; Wanner, Gerhard (1993), Solving ordinary differential equations I: Nonstiff problems (§III.10), Berlin, New York: Springer-Verlag.
\bibitem{Caruso} F. Caruso, V. Oguri, Revista Brasileira de Ensino de Fisica, v. \textbf{36}, n. 2, 2310 (2014).
\bibitem{feynman} We follow the perturbation strategy found in R. P. Feynman, ``Statistical Mechanics-A Set of Lectures'', Westview Press (1998);
\end{thebibliography}
\end{document}